# Factor Machine: Mixed-signal Architecture for Fine-Grained Graph-Based Computing

Piotr Dudek

Department of EEE, The University of Manchester

**Abstract**

This paper proposes the design and implementation strategy of a novel computing architecture, the Factor Machine. The work is a step towards a general-purpose parallel system operating in a non-sequential manner, exploiting processing/memory co-integration and replacing the traditional Turing/von Neumann model of a computer system with a framework based on "factorised computation". This architecture is inspired by neural information processing principles and aims to progress the development of brain-like machine intelligence systems, through providing a computing substrate designed from the ground up to enable efficient implementations of algorithms based on relational networks. The paper provides a rationale for such machine, in the context of the history of computing, and more recent developments in neuromorphic hardware, reviews its general features, and proposes a mixed-signal hardware implementation, based on using analogue circuits to carry out computation and localised and sparse communication between the compute units.

## 1. Introduction

The progress of electronic computing over the past 70 years has been fuelled by the spectacular advances in microelectronics fabrication technology, as described by the Moore's Law [1], and the continuing appeal and applicability of the idea of universal general-purpose computation based on the automated computing machinery, known as the Turing Machine [2]. The concept of computation as activity that can be carried out by a sequential machine, following step-wise instructions, and fetching/storing intermediate results of these operations in a memory bank (or a tape in the Turing Machine) is central to the design of modern computers. The hardware scheme for implementing this stored-program automated computation machinery is nowadays often commonly termed "von Neumann architecture", following the original developments of digital computing systems [3]. Its success and prevalence is in large part due to a direct correspondence between this architecture, and the algorithmic way of thinking which typically characterises human way of understanding computation. But, while it has been enormously successful, its limits are also becoming increasingly apparent.

In a practical hardware implementation, the processor (CPU) is physically distant from the memory. The CPU-memory communications take time and energy, and are the most critical component responsible for the speed and power consumption of contemporary computer systems. The innovations in memory systems (cache hierarchies etc) have produced improvements, but fundamentally this limitation, that has become known as the "von Neumann bottleneck", is always there. Much research on computer architecture has been therefore recently directed into alleviating this issue, through closer integration of processing and memory circuits, and it is clear that such integration is necessary to achieve improvements in performance and energy efficiency of computing systems.

Second limitation of the "von Neumann" architecture is the single-threaded performance of the sequential computation, ultimately limiting the speed of the computations given feasible clock frequencies. Innovations in CPU architectures, including pipelining, multi-threading, multiple issue, SIMD, etc. have brought improvements [4]. Multi-core architectures - allowing threads to be executed concurrently on separate processors - are now prevalent, providing the execution speed and power efficiency improvements at the cost of additional hardware. The parallelisation allows us to keep increasing computational performance, and systems such as GPU's [5] or some of the recent many-core machines [6] include thousands of processing cores working in parallel. Still, essentially these are assemblies of von Neumann machines. A recent trend is that of an addition of dedicated hardware accelerators, in particular to address the increasing demands of so-called "AI" computations, which are typically artificial neural networks [7,8] . Generally, these contain circuitry to organise an efficient dataflow through the large number of compute units, such as multiply and add circuits (as these are the most plentiful arithmetic operations at the core of the current AI algorithms). The concept of the step-wise computation remains unchanged, but the operations are sequenced in space as well as time, to allow more parallelism. There is much to be gained optimising hardware for these operations, perhaps even using new device technologies [9]. Optimised arithmetic units can enable efficient and parallel hardware implementations. However, these accelerators are typically embedded in a sequential (von-Neumann) programmable computer architecture that provides the general-purpose facility to the overall system.

So, while "improving von-Neumann" has been the goal of many recent endeavours, the Turing Machine concept of computation as a sequence of steps dictated by a program is hardly ever challenged. Yet, there should be alternatives. Some of these are hinted to by the developments in neuromorphic computing, the idea of constructing computing systems more closely resembling the brains. It is clear that, while some of human thinking is very much that of following algorithmic, step-wise procedures, most of it is not. The brains are solving problems through massively parallel distributed dynamical form of computation, where local computing units (e.g. neurons or neural populations), operating on local memories embedded therein (e.g. synapses), somehow collectively achieve a coherent state that results in intelligent behaviour. Attempts have been made at capturing these kinds of computations through abstractions such as Spiking Neural Network models, including notable neuromorphic hardware developments in this field in academia and industry (SpiNNaker [10], Intel Loihi [11], IBM True North [12], etc), but so far with limited success in terms of providing competitive solutions, let alone a general-purpose computing substrate, or one that would provide a practical vehicle for implementing intelligent machines with capabilities and performance comparable with the human brain. It might be that this bottom-up approach, taking a spiking neuron as the fundamental compute unit, starts too far down in the mechanics of biological implementation to make it possible to unravel the computational essence of the human intelligence. Perhaps a different level of abstraction, a different model of computation, is needed to bridge the gap between the brains and feasible novel general-purpose hardware architectures.

## 2. Factorised computation

We should be interested in exploring the design of machines based on a computational frameworks alternative to the Turing machine model, but still providing general-purpose functionality and applicability to a wide class of problems (rather than acceleration of a particular parallel model e.g. a neural network etc.). A promising attempt at such framework is provided by the idea of *Factorised Computation* outlined in [13]. It attempts to identify some general principles for constructing massively parallel, distributed computing systems, whose programming is based on the declaration of quantities of interest and their relations. It also addresses mechanisms by which a problem stated in such way can be solved by a concurrent system. This framework is rather generic, but in a possible instantiation of this concept, we can imagine a system consisting of a set of variables (that can be inputs, outputs and latent system variables), and a set of relations between these variables (that can be represented for instance as functions or constraints). A bipartite graph, with certain nodes representing variables, others representing their relations, provides a clear graphical representation of such system (see a simple example in Figure 1). The "programming" of the system consists in defining the relations between the variables and specifying those quantities that are known. Note that all these relationships are local, and each system component "sees" only itself, and the components it is connected to. The execution of the system proceeds via exchange of local information (messages) between the nodes, alongside the edges in the graph. The local computations at the nodes resolve information carried by these messages (which may be conflicting or incomplete) in such a way that, through the local computations and interactions, a global consensus is achieved about the state of the system variables. This steady-state consensus is therefore a solution to the computational problem. Or, in a more general way, and more consistent with the idea of building artificial intelligent systems, there is no steady-state solution, but instead, in response to the changing inputs, the system dynamics progress in some desirable way.

Overall, the approach to solving problems using a form of factorised computation is not new, and a variety of models/algorithms have been expressed in a way compatible with this framework. Many of those have clear and important applications. For instance, algorithms for Belief Propagation in Bayesian Networks can be solved through message passing in graphs representing probabilistic networks [14]. One notable example is that of sum-product algorithm for factor graphs, used for decoding Turbo Codes and LPDC codes at the core of modern communications systems [15]. In general, many optimisation problems can often be naturally expressed in this factorised/graph-based way. When variables are operating over discrete domains, such system can express constraint satisfaction problems. There are many other algorithms and approaches that use this form of computation. Even systems such as spiking neural networks, or brain models based on dynamic neural fields, can be seen as a special cases of such system. The Factorised Computation framework [13] attempts to gain insights into properties of these systems as a form of general-purpose/universal parallel computation, develop new techniques for constructing them, and investigate how these can be applied to a variety of problems, including the understanding of biological computing systems.

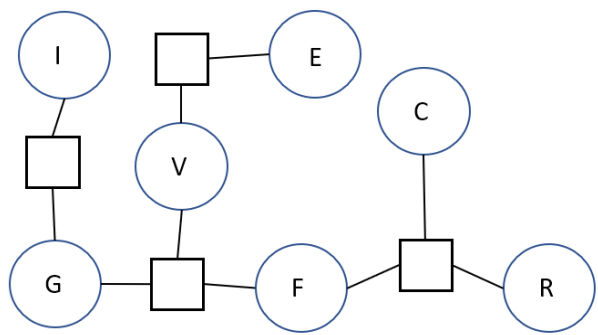

*Fig.1. Example of a graph representing a "factorised" system. Circle nodes are variables (in this case, scalars, vectors or arrays), square nodes are relations between them. This example, from [16], represents a system for jointly inferring global rotation R, image I, and optic flow F from a stream of event-camera data E, and camera calibration model C, using latent variables for temporal contrast V and gradient map G.*

## 3. Factor Machine

What is of interest to us, as hardware engineers, is the idea that the factorised computation way of thinking about parallel distributed systems can not only offer new insights into the idea of computation, but that it can also provide a blueprint for the construction of new intelligent machines. We shall think of this as a unique form of computation (not merely as the vehicle for problem representation to be later implemented as an algorithm on a Turing-style machine) and therefore explore a way of implementing hardware machinery designed from ground-up for this computation style. At the same time, we are suggesting that early attempts at building such new type of computing machinery, no matter how imperfect they may initially be, can bootstrap this research direction, providing a substrate where exercises of mapping problem onto such machines can be carried out, ultimately stimulating interest and development of this form of computing.

The fine-grained parallel systems, consisting of a very large number of simple computing units are undoubtedly of interest in terms of future energy efficient computing systems. And, which is important from a practical development point of view, a design deriving the power from the quantity and interaction of simple units (as opposed to fewer, more complex units, or a heterogenous system) is amenable to the Very Large Scale of Integration (VLSI) design methodologies, implementable in today's CMOS silicon integrated circuit technologies, and manageable as a proof-of-concept design by a small research team.

Given that the concepts underlying the factorised computation idea are very general, the initial attempts at building a general-purpose computers based on these concepts will need to be based on some intuitive decisions. There is a vast space of possible system architectures, hardware designs and implementation technologies that could be explored. One direction, motivated in part by the opportunity to break free from associating general-purpose computation with digital logic, would be to explore building such systems in analogue and mixed-signal hardware. Analogue computing is promising in terms of both speed and energy efficiency (if a level of noise and limited accuracy is acceptable) [17,18]. Therefore, without prejudice to other forms of implementation (in particular using conventional digital logic circuitry) we think this direction could be the most interesting to explore. But, since analogue design is a difficult and time-consuming art, and success is a hard-fought result of the design effort, we would also advocate using digital implementations for more agile design space explorations. The digital architecture ideas can be also more reliably evaluated via simulations and can provide useful comparisons with the analogue approach.

The starting idea for the kind of Factor Machine we might want to build is outlined below. The machine consists of a very large number of identical cells. Each cell in the system combines several variable and relation nodes of a computational graph. Each cell, as shown in Figure 2#, comprises storage elements that hold the variables (or components of variable representation, for instance in case of stochastic variables), and a configurable processing circuitry, defining the relations between the variables and implementing the required computations. The processors have a set of functional components, for arithmetic operations such as multiplication, addition, etc., and perhaps some specialised operations (e.g. max, or winner-take-all). The processors implement the required relations and operate on local variables. We envisage this as a discrete-time system, where the updates can occur only when necessary (i.e. in the response of to the *event* of the variable update). In the purest from, the processing circuitry implements the relationships in a dataflow manner, but we may consider the implementation of a small sequential state-machine control subsystem, for efficient utilisation of hardware resources (e.g. a multiplier cell could be used twice during one variable update). The cells are interconnected via a configurable routing fabric, that is be used to transmit the messages between the cells, and effectively define the structure of the computational graph.

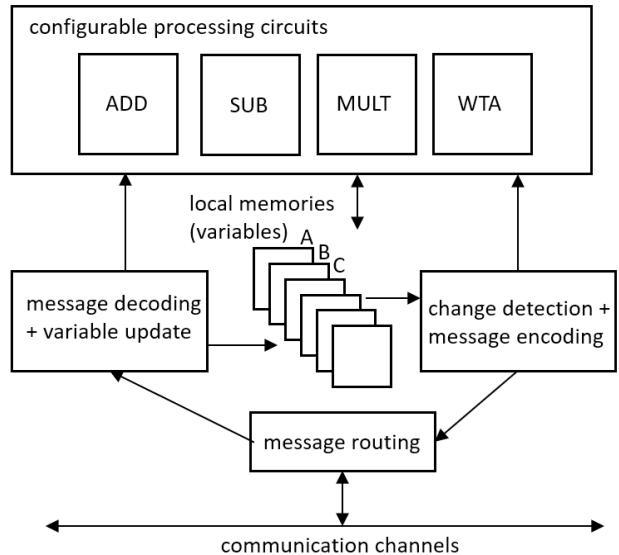

*Fig.2. A sketch of the basic cell in the Factor Machine. The cell implements nodes (variables + relations) of the graph.*

One of the key features enabling efficient computing here is the co-location of memory and processing circuitry. The basic cells are relatively simple, and thousands of these are replicated over the surface of the chip. The mapping of the computing graph onto the hardware cell array should ensure the communications are local, but in a flexible general-purpose architecture some longer connections are unavoidable. To avoid fetching data from distant locations, each cell should contain the copies of the variables of interest (belonging to the node, as well as the logically neighbouring nodes). In the probabilistic interpretation, these represent local "beliefs" about the state of these variables. In a more general computing or optimisation framework, these simply represent the locally stored approximation to the true value of the variable.

In the analogue system, the variables would be stored in analogue memories, as charge on capacitors, and messages would most likely be simply digital pulses, indicating the discrete change/update of the variable (similar, perhaps, to a sigma-delta modulation). In the digital versions of this architecture, there could be a variety of messages and variable representations, from sending multi-bit values directly over the communication network, to representing stochastic variables as probability density functions, and using probabilistic methods such as Gibbs sampling to evolve the state of the system. In any case, the routing system should work with dedicated variable update and change detection circuitry in each cell, to handle the variable update and communication mechanisms.

This architecture has the required features of a general-purpose 'factorised computation' engine. The system would allow a construction of a variety of computational graphs, by flexibly routing the messages between the cells. A number of strategies for variable update, and configurable computation in each cell could be explored. Therefore a variety of "applications" could be demonstrated with this machine, for instance as a solver for a variety of constraint satisfaction problems (e.g. graph colouring, Sudoku), implementation of Belief Propagation in Factor Graphs, implementation of a Spiking Neural Network, implementation of a sensory fusion system (e.g. combining visual odometry and inertial odometry sensors), implementation of a controller for a de-centralised robotic manipulator system, etc.

It is well known that many interesting and high-performing hardware architectures have failed, not because of the deficient hardware technology, but because of software (or rather the lack of it, or the difficulty in using it). It would be therefore essential that a significant effort is put into the development of software tools. It is (somewhat wishfully) thought, that the very existence of the hardware machine would stimulate and bootstrap the work on mapping various problem onto such machines, as well as the theoretical underpinnings of the factorised computation scheme.

Finally, it should be obvious that the overall processing scheme of this machine could be emulated on a general parallel processor hardware. Especially parallel systems implementing programmable processor nodes and efficient lightweight communication fabric [10,19,29] would be particularly suited. The key distinguishing features of the proposed Factor Machine hardware are that the system is build purposely to execute the factorised computation scheme, the processing cells are very simple, allowing thousands, eventually millions of these to be implemented on a single chip, and there is intimate memory-processing colocation, promising low-power operation. The hardware in each cell has circuitry dedicated to specific key operations such as sparse variable update, and is build ground-up for efficiency of the fine-grained computation (including, for instance, event-triggered computation). The potential for using analogue computation also distinguishes the proposed device from the majority of current parallel hardware computing systems.

## 4. Conclusions

This paper attempts to signpost a research direction, which explores an alternative to both conventional parallel systems, based on assemblies of concurrently operating sequential von-Neumann-type machines, and dataflow hardware architectures, such as neural network accelerators, crafted to speed up some common arithmetic/logic operations. We propose the idea of a Factor Machine, a novel computer architecture based on a large assembly of fine-grained compute elements, incorporating local memory and processing circuits. The architecture is based on theoretical underpinnings of factorised computation, and is intended to provide a practical alternative to a sequential, Turing machine mode of computation. The implementation of such architecture is possible using digital as well as analogue circuit technologies. We postulated that the properties of this architecture, including processing/memory integration, massive parallelism, distributed and de-centralised computation, make it suitable for the future generations of energy efficient computing systems, in particular, systems implementing artificial intelligence (and not necessarily based on a connectionist neural network approaches). There is a vast design space of possible implementations based on this framework, and much opportunity for exploring the alternatives. The proposed system has its roots in the ideas of neuromorphic computing, but provides an alternative to the prevailing trends for implementing spiking neural network hardware or focussing on materials and devices for dense realisations of synaptic arrays. It might help stimulate interest in addressing higher levels of abstraction when thinking about brain-inspired hardware systems. It might also stimulate a broader interest in theories of computation. Novel computing paradigms could be investigated through simulations, but there is much merit in constructing hardware to aid this effort. The first designs for electronic computing were influenced by the theoretical idea of Turing's sequential state-machine. In turn, the computing machinery itself has enabled and inspired the conceptual developments of more sophisticated sequential computing ideas. The new computational ideas need to be constrained, and perhaps inspired by, their physical instantiations.

## Acknowledgements

The ideas outlined in this paper have originated from discussions with Matthew Cook, at INI Zurich, Telluride, CapoCaccia, and other places. This paper is dedicated to Steve Furber, whose insights and work on computer architecture, neuromorphic computing, and large-scale hardware for simulating brains have been always inspiring.